\documentclass[preprint,showpacs,preprintnumbers,amsmath,amssymb]{revtex4}
\usepackage{booktabs}
\usepackage{mathrsfs}
\usepackage{epsfig}
\usepackage{graphicx}
\usepackage{dcolumn}
\usepackage{bm}
\usepackage{amsmath}
\usepackage{slashed}
\usepackage{multirow}

\let\jnfont=\rm
\def\NPB#1,{{\jnfont Nucl.\ Phys.\ B }{\bf #1},}
\def\PLB#1,{{\jnfont Phys.\ Lett.\ B }{\bf #1},}
\def\EPJC#1,{{\jnfont Eur.\ Phys.\ Jour.\ C }{\bf #1},}
\def\PRD#1,{{\jnfont Phys.\ Rev.\ D }{\bf #1},}
\def\PRL#1,{{\jnfont Phys.\ Rev.\ Lett.\ }{\bf #1},}
\def\MPLA#1,{{\jnfont Mod.\ Phys.\ Lett.\ A }{\bf #1},}
\def\JPG#1,{{\jnfont J.\ Phys.\ G}{\bf #1},}
\def\CTP#1,{{\jnfont Commun.\ Theor.\ Phys.\ }{\bf #1},}
\def\ZPC#1,{{\jnfont Z.\ Phys.\ C }{\bf #1},}
\def\JHEP#1,{{\jnfont JHEP \ }{\bf #1},}
\def\Rv{\not{\hbox{\kern-1pt $R$}}}
\def\p{\not{\hbox{\kern-3pt $p$}}}

\newcommand{\met}{\not\!\!\!E_{T}}

\begin{document}

\title{Associated production of single top and Higgs at the LHC in the littlest Higgs
model with T-parity}

\author{ Bingfang Yang$^{1,2}$}
\author{Jinzhong Han$^{3}$}
\author{ Ning Liu$^{1}$}
\affiliation{$^1$ College of Physics $\&$ Electronic Engineering,
Henan Normal University, Xinxiang 453007, China\\
$^2$ School of Materials Science and Engineering, Henan Polytechnic
University, Jiaozuo 454000, China\\
$^3$Department of Physics and Electronic Engineering, Zhoukou Normal
University, Zhoukou, 466001, China
   \vspace*{1.5cm} }%

\date{\today}

\begin{abstract}

In the littlest Higgs model with T-parity (LHT), we study the
$t$-channel single top production in association with a Higgs boson
at 8 and 14 TeV LHC. We find that the cross section can be enhanced
obviously in this model compared to the Standard Model. By
performing a simple parton-level simulation through $pp\rightarrow
t(\rightarrow \ell^{+}\nu b)h(\rightarrow b\bar{b})j$ at 14 TeV LHC,
we find that the observability of the signal is promising in the
favorable parameter space.

\end{abstract}
 \maketitle

\section{INTRODUCTION}

On the 4th of July 2012, the ATLAS and CMS collaborations at the
Large Hadron Collider (LHC) have discovered a Higgs-like resonance
about 125 GeV\cite{higgs-lhc}. With current data, all properties of
the discovered Higgs boson turn out to be in rough agreement with
expectations of the Standard Model (SM)\cite{LHC-higgs}. Due to the
large experimental uncertainties, there remains a plenty of room for
new physics in Higgs sector\cite{higgs-coupling}. In order to
ultimately establish its nature, a precise measurement of the Higgs
couplings is essential and this task will be performed in the next
phase of the LHC and future Higgs factory.

The Yukawa couplings play an important role in probing the new
physics since they are sensitive to new flavor dynamics. In view of
the large mass, the top quark owns the strongest Yukawa coupling so
that it is an appropriate probe for the electroweak symmetry
breaking (EWSB) mechanism and new physics\cite{top}. As a direct
probe of the top Yukawa coupling, the production of a top pair
associated with a Higgs boson($t\bar{t}h$ production) is a golden
channel and has received great attention by the
experimenters\cite{tth-exp} and theorists\cite{tth-theory}. However,
the information on the relative sign between the top Yukawa coupling
and Higgs coupling to gauge bosons will still be lacking. In this
respect, the production of a single top quark associated with a
Higgs boson ($thj$ production) can bring a rather unique
possibility\cite{thj-theory}. The $pp\rightarrow thj$ production
process can be divided into three different modes characterised by
the virtuality of the $W$ boson\cite{thj-modes}: (i) $t$-channel,
where the $W$ is spacelike; (ii) $s$-channel, where the $W$ is
timelike; (iii) $W$-associated, where there is emission of a real
$W$ boson. Besides, the anomalous $pp\rightarrow thj$ production
process can be induced by the top-Higgs flavor changing neutral
current(FCNC) interactions\cite{thj-work}.

The littlest Higgs model with T-parity (LHT)\cite{LHT} was proposed
as a possible solution to the hierarchy problem and so far remains a
popular candidate of new physics. The LHT model predicts new gauge
bosons, scalars, mirror fermions and top partner, where the T-even
top partner $T_{+}$ can contribute to the $pp\rightarrow thj$
process. Furthermore, some Higgs couplings are modified at the high
order and this effect can also influence the process $pp\rightarrow
thj$. By performing the detailed analysis on the process
$pp\rightarrow thj$ may provide a good opportunity to probe the LHT
signal. At the LHC, the $t$-channel process dominates amongst these
production modes and the related work has been studied in the
littlest Higgs model\cite{LHthj}. In this work, we focus on
$t$-channel process and investigate the observability of
$pp\rightarrow thj$ with sequent decays $t\rightarrow \ell^{+}\nu b$
and $h\rightarrow b\bar{b}$ at 14 TeV LHC in the LHT model.

The paper is organized as follows. In Sec.II we give a brief review
of the LHT model related to our work. In Sec.III we calculate the
$t$-channel process of $pp\rightarrow thj$ at the LHC and explore
the observability by performing a parton-level simulation. Finally,
we give a summary in Sec.IV.

\section{A brief review of the LHT model}

The LHT model was based on a non-linear $\sigma$ model describing an
$SU(5)/SO(5)$ symmetry breaking, with the global group $SU(5)$ being
spontaneously broken into $SO(5)$ by a $5\times5$ symmetric tensor
at the scale $f\sim \mathcal O$(TeV).

In the top Yukawa Sector, in order to cancel the large radiative
correction to Higgs mass parameter induced by top quark, an
additional top partner $T_{+}$ is introduced, which is even under
T-parity and transforms as a singlet under $SU(2)_{L}$. The
implementation of T-parity requires a T-odd mirror partner $T_{-}$.
For the top Yukawa interaction, one can write down the following
$SU(5)$ and T-parity invariant Lagrangian\cite{LHT}:
\begin{eqnarray}
{\cal L}_t&=& -\frac{\lambda_1f}{2\sqrt{2}} \epsilon_{ijk}
\epsilon_{xy} \left[(\bar{Q}_1)_i \Sigma_{jx} \Sigma_{ky}-
(\bar{Q}_2 \Sigma_0)_i \tilde{\Sigma}_{jx} \tilde{\Sigma}_{ky}
\right] u_{R_+}
\nonumber \\
&& -\lambda_2 f (\bar{U}_{L_1} U_{R_1}+\bar{U}_{L_2} U_{R_2}) +{\rm
h.c.}, \label{top_yukawa_int}
\end{eqnarray}
where $\epsilon_{ijk}$ and $\epsilon_{xy}$ are antisymmetric
tensors, and $i,~j$ and $k$ run over $1,2,3$ and $x$ and $y$ over
$4,5$. $u_{R_+}$ and $U_{R_i}$ $(i=1,2)$ are $SU(2)$ singlets.

The heavy quark $T_{+}$ mix with the SM top-quark and leads to a
modification of the top quark couplings relatively to the SM. The
mixing can be parameterized by dimensionless ratio
$R=\lambda_1/\lambda_2$, where $\lambda_1$ and $\lambda_2$ are two
dimensionless top quark Yukawa couplings. This mixing parameter can
also be used by $x_{L}$ with
\begin{equation}
x_{L}=\frac{R^{2}}{1+R^{2}}
\end{equation}

Their masses up to $\mathcal O(v^{2}/f^{2})$ are given by
\begin{eqnarray}
&&m_t=\lambda_2 \sqrt{x_L }v \left[ 1 + \frac{v^2}{f^2} \left(
-\frac{1}{3} + \frac{1}{2} x_L \left( 1 - x_L \right) \right)
\right]\\
&&m_{T_{+}}=\frac{f}{v}\frac{m_{t}}{\sqrt{x_{L}(1-x_{L})}}\left[1+\frac{v^{2}}{f^{2}}\left(\frac{1}{3}-x_{L}(1-x_{L})\right)\right]\\
&&m_{T_{-}}=\frac{f}{v}\frac{m_{t}}{\sqrt{x_{L}}}\left[1+\frac{v^{2}}{f^{2}}\left(\frac{1}{3}-\frac{1}{2}x_{L}(1-x_{L})\right)\right]
\end{eqnarray}
where $v=v_{SM}(1+\frac{1}{12}\frac{v_{SM}^2}{f^2})$ and $v_{SM}=
246$ GeV is the SM Higgs VEV. Some typical Higgs couplings involved
in our calculations are given by
\begin{eqnarray}
&&V_{Hb\bar{b}}=-\frac{m_{b}}{v}\left(1-\frac{1}{6}\frac{v^2}{f^2}\right),\\
&&V_{HW_{\mu}W_{\nu}}=\frac{2m_{W}^{2}}{v}\left(1-\frac{1}{6}\frac{v^2}{f^2}\right)g_{\mu\nu},\\
&&V_{W_{\mu}\bar{t}b}=\frac{V_{tb}}{\sqrt{2}}g\gamma_{\mu}\left(1-\frac{x_{L}^{2}}{2}\frac{v^{2}}{f^{2}}\right)P_{L},\\
&&V_{W_{\mu}\bar{T}_{+}b}=\frac{V_{tb}}{\sqrt{2}}g\gamma_{\mu}x_{L}\frac{v}{f}P_{L},\\
&&V_{ht\bar{t}}=-\frac{m_t}{v}\left[1+\frac{v^2}{f^2}(-\frac{2}{3}+x_{L}-x_{L}^{2})\right],\\
&&V_{ht\bar{T}_{+}}=-m_{t}\left[\frac{(1-x_{L})}{f}P_{R}-\frac{\sqrt{x_{L}}}{v\sqrt{1-x_{L}}}P_{L}\right].
\end{eqnarray}
where $P_{L}=\frac{1-\gamma_{5}}{2}$ and
$P_{R}=\frac{1+\gamma_{5}}{2}$ are chirality projection operators.
The Higgs coupling with down-type quarks have two different
cases\cite{case}, namely case A and case B.

\section{Numerical results and discussions}
\begin{figure}[htbp]
\scalebox{0.55}{\epsfig{file=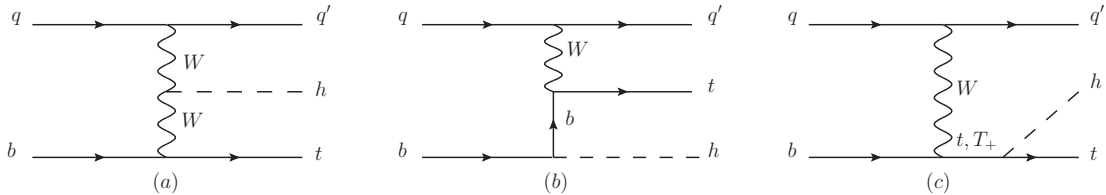}}\vspace{-0.5cm}\caption{
Feynman diagrams for $pp \to thj$ in the LHT model at the tree
level.}\label{thjlht}
\end{figure}

In the LHT model, the lowest-order Feynman diagrams of the process
$pp \to thj(j\neq b)$ are shown in Fig.\ref{thjlht}. We can see that
the T-even heavy quark $T_{+}$ contributes this process through the
Fig.\ref{thjlht}(c). In our calculations, the conjugate process $pp
\to \bar{t}hj$ has been considered, unless otherwise noted.

We compute the cross sections using the $\textsf{MadGraph
5}$\cite{mad5} (and checked against those obtained by
$\textsf{CalcHEP 3.6.22}$\cite{calchep}), where the CTEQ6L
\cite{cteq} is used as the parton distribution function and the
renormalization scale $\mu_R$ and factorization scale $\mu_F$ is set
to be $\mu_R=\mu_F=m_{t}+m_{h}$. The relevant SM input parameters
are taken as follows \cite{pdg}:
\begin{align}
m_t = 173.07{\rm ~GeV},\quad &m_{Z} =91.1876 {\rm ~GeV}, \quad
\alpha(m_Z) = 1/128, \\ \nonumber \sin^{2}\theta_W = 0.231,\quad
&m_h =125 {\rm ~GeV}, \quad \alpha_{s}(m_Z)=0.1185.
\end{align}

The relative correction of the cross section can be defined as
\begin{equation}
\delta \sigma/\sigma=\frac{\sigma-\sigma_{SM}}{\sigma_{SM}}.
\end{equation}
In our calculations, the leading-order cross sections for the
processes $pp \to thj$ in the SM are taken as $\sigma_{SM}^{8\rm
TeV}=16.4$fb and $\sigma_{SM}^{14\rm TeV}=80.4$fb.

\begin{figure}[htbp]
\scalebox{0.35}{\epsfig{file=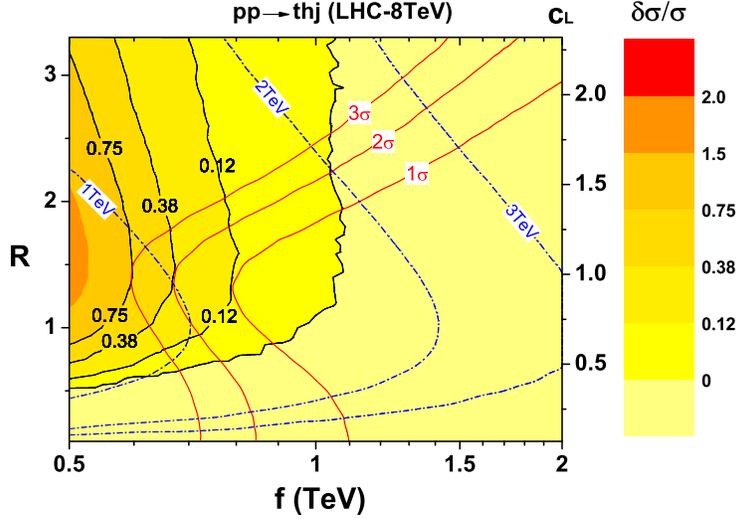}} \vspace{-0.5cm}\caption{The
relative corrections $(\delta \sigma/\sigma)_{thj}$ at 8 TeV LHC in
the LHT model. The red solid lines respectively represent the
$1\sigma$, $2\sigma$ and $3\sigma$ confidence regions, the blue
dash-dot lines respectively represent the $m_{T_{+}}=1$ TeV, 2 TeV
and 3 TeV.}\label{dcross8}
\end{figure}

\begin{figure}[htbp]
\scalebox{0.35}{\epsfig{file=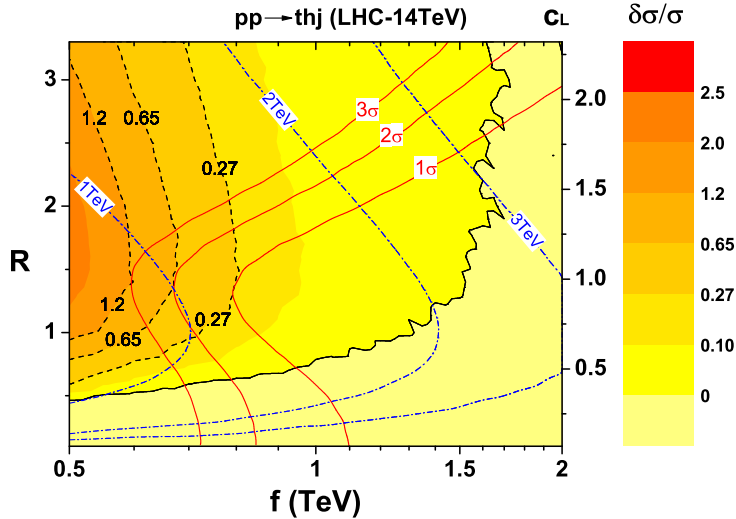}}\vspace{-0.5cm}\caption{The
relative corrections $(\delta \sigma/\sigma)_{thj}$ at 14 TeV LHC in
the LHT model. The red solid lines respectively represent the
$1\sigma$, $2\sigma$ and $3\sigma$ confidence regions, the blue
dash-dot lines respectively represent the $m_{T_{+}}=1$ TeV, 2 TeV
and 3 TeV.}\label{dcross14}
\end{figure}

Our results show that the features of the process $pp \to thj$ are
very similar for the case A and case B, so here we focus on the case
A scenario. The LHT parameters related to our calculations are the
scale $f$ and the ratio $R$. Considering the consistent constraints
in Refs.\cite{constraint}, we require the scale $f$ and the ratio
$R$ to vary in the range $500$ GeV$\leq f\leq 2000$ GeV and $0.1
\leq R\leq 3.3$. Combined with the global fit of the current Higgs
data and the electroweak precision observables(EWPO) in
Ref.\cite{fit}, the confidence regions (corresponding to $1\sigma$,
$2\sigma$ and $3\sigma$ ranges for case A) are provided in
Figs.(\ref{dcross8},\ref{dcross14}). Furthermore, according to
Refs.\cite{DM-LHT}, we can see that the constraints on the LHT
parameters from the dark matter observations are weaker than the
Higgs data and EWPO at $2\sigma$ level, which means that the
parameter space allowed by the Higgs data and EWPO can also satisfy
the dark matter constraints.

In Fig.\ref{dcross8} and Fig.\ref{dcross14}, we show the relative
corrections $\delta \sigma/\sigma$ of the processes $pp \to thj$ at
the 8 and 14 TeV LHC in the LHT model, respectively. From the
Fig.\ref{dcross8} and Fig.\ref{dcross14}, we can see that the
relative corrections $\delta \sigma/\sigma$ of $pp \to thj$ at 8 and
14 TeV LHC can be respectively reach 38\% and 65\% at 2$\sigma$
level. We find that these large corrections mainly come from the
resonance decay of the heavy quark $T_{+}$ in the
Fig.\ref{thjlht}(c). Furthermore, we can see that the relative
corrections $\delta \sigma/\sigma$ are negative in considerable
regions for 8 TeV LHC and non-negligible regions for 14 TeV LHC. The
main reasons are as follows:

Due to the small coupling $hb\bar{b}$, the main contribution to the
$pp \to thj$ comes from Fig.\ref{thjlht}(a,c). If we take no account
of the heavy quark $T_{+}$, we can see that the couplings $hWW$ and
$ht\bar{t}$ have the opposite sign so that the contributions of
Fig.\ref{thjlht}(a) and Fig.\ref{thjlht}(c) cancel each other.
According to the Eq.(10), we can see that the left-handed part
($c_{L}=m_{t}\frac{R}{v}$) of the coupling $ht\bar{T}_{+}$ has the
same sign as the coupling $hWW$ so that their contributions enhance
each other. The same thing happens between the right-handed part
($c_{R}=-m_{t}\frac{(1-x_{L})}{f}$) of the coupling $ht\bar{T}_{+}$
and the coupling $ht\bar{t}$. As a result, the total contribution
induced by the top partner $T_{+}$ depends on the surplus after the
cancelation between $c_{L}$ contribution and $c_{R}$ contribution.
One can notice that the left-handed coupling $c_{L}(\propto R)$ is
proportional to the ratio $R$ and dominates the contribution from
the heavy quark $T_{+}$. Moreover, the Higgs couplings in the LHT
model are modified at $\mathcal O(v^{2}/f^{2})$, which can decrease
the $thj$ cross section. Combining these factors above, we can see
that the large relative corrections $\delta \sigma/\sigma$ come from
the region that has small $f$, small $m_{T_{+}}$ and large $c_{L}$.
By contrast, the small or negative relative corrections $\delta
\sigma/\sigma$ come from the region that has large $f$, large
$m_{T_{+}}$, small $c_{L}$ or the combination of them. Due to the
lower centre-of-mass energy, the relative corrections $\delta
\sigma/\sigma$ at 8 TeV LHC are suppressed by the large $m_{T_{+}}$
more strongly so that the negative $\delta\sigma/\sigma$ regions are
larger compared to the case for 14 TeV LHC. Furthermore, for the
same ratio $R$, we can see that the relative corrections $\delta
\sigma/\sigma$ of $pp \to thj$ at 8 and 14 TeV LHC both decrease
with the scale $f$ increasing, which means that the LHT effect
decouples with the scale $f$ increasing.

In the following calculations, we will perform a simple parton-level
simulation and explore the sensitivity of 14 TeV LHC through the
channel $pp\rightarrow t(\rightarrow \ell^{+}\nu b)h(\rightarrow
b\bar{b})j$, the signal is characterised by
\begin{equation}
1 {\rm forward~jet} + 3 b + \ell^{+} + \met
\end{equation}
where $j$ denotes the light jets and $\ell=e,\mu$. The most relevant
backgrounds can be divided into two classes:

(i) reducible backgrounds, $pp \to t\bar{t}(\to \bar{b}\bar{c}s)$
and $pp \to t\bar{t}(\to \bar{b}\bar{c}s)j$;

(ii) irreducible backgrounds, $pp \to tZ(\to b\bar{b})j$ and $pp \to
tb\bar{b}j$.

Signal and background events have been generated at the parton level
using $\textsf{MadGraph 5}$, the subsequent simulations are
performed by $\textsf{MadAnalysis 5}$\cite{madanalysis}. To simulate
$b$-tagging, we take moderate single $b$-tagging efficiency
$\epsilon_{b}=0.6$ for $b$-jets in the final state. We also include
charm mistag probability $\epsilon_{c}=0.08$ and light jet mistag
probability $\epsilon_{j}=0.004$ in the reducible
backgrounds\cite{b-tag}. Follow the analysis on $t\bar{t}h$
signature by ATLAS and CMS collaborations\cite{tth-exp} at the LHC
Run-I, we chose the basic cuts as follows:
\begin{eqnarray}\label{basic}
\nonumber&&\Delta R_{ij} >  0.4\ ,\quad  i,j = b,j \ \text{or}\  \ell \\
&& p_{T}^b > 25 \ \text{GeV},\quad |\eta_b| <2.5 \\
\nonumber && p_{T}^\ell > 25 \ \text{GeV}, \quad  |\eta_\ell|<2.5  \\
\nonumber && p_{T}^j > 25 \ \text{GeV},\quad  |\eta_j|<5.
\end{eqnarray}

\begin{figure}[htbp]
\scalebox{0.31}{\epsfig{file=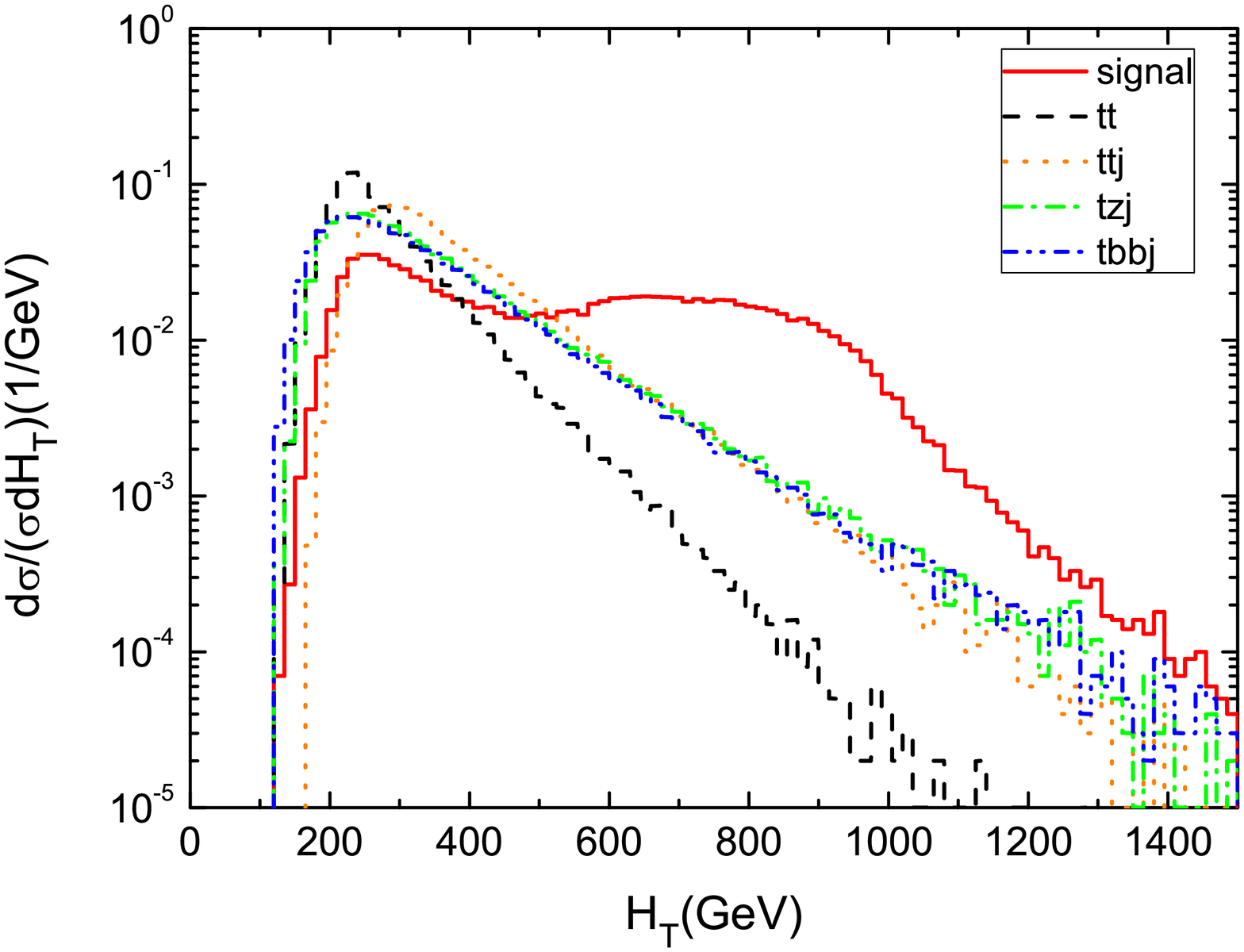}}\hspace{-1.2cm}
\scalebox{0.31}{\epsfig{file=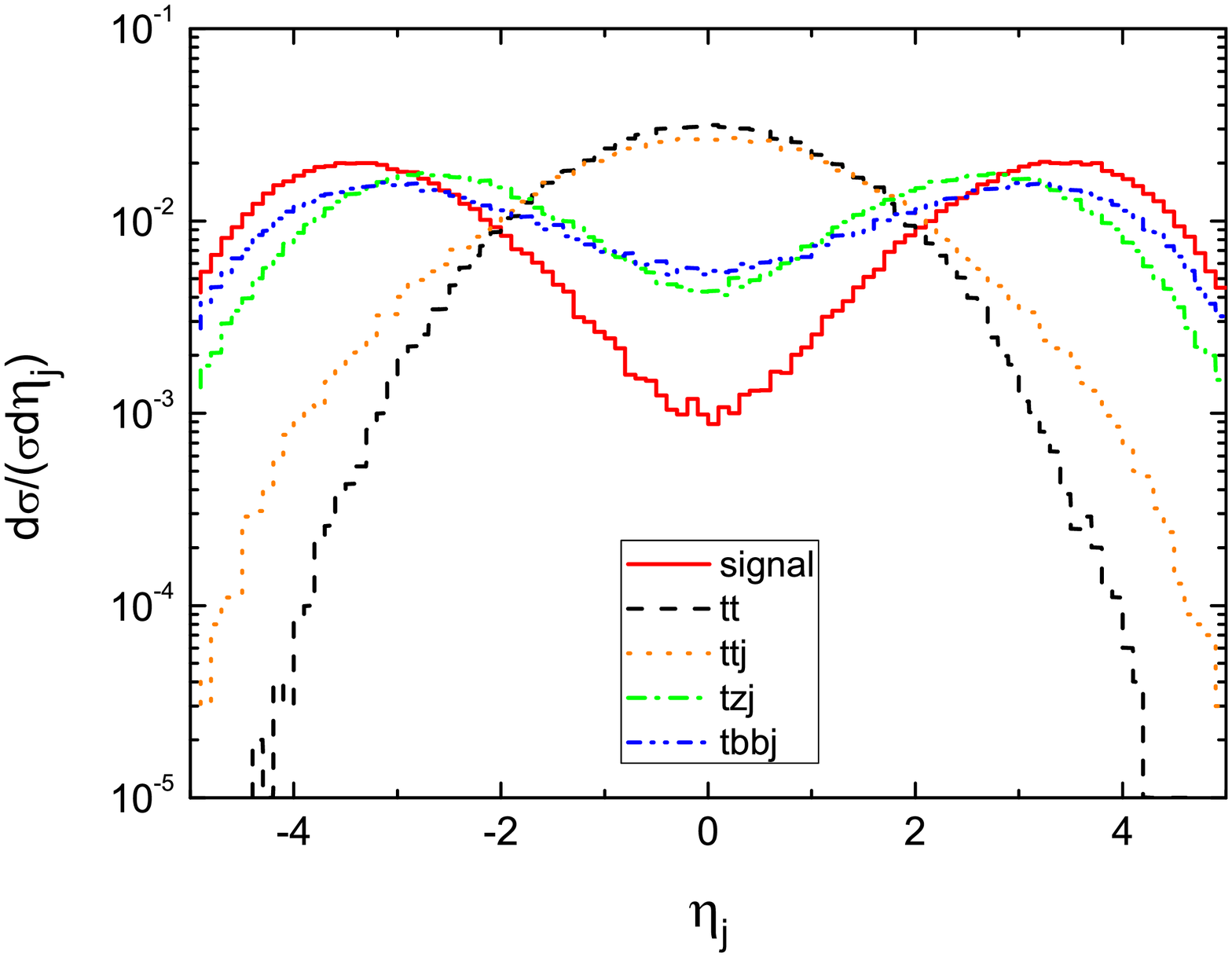}}\vspace{-0.4cm}
\scalebox{0.31}{\epsfig{file=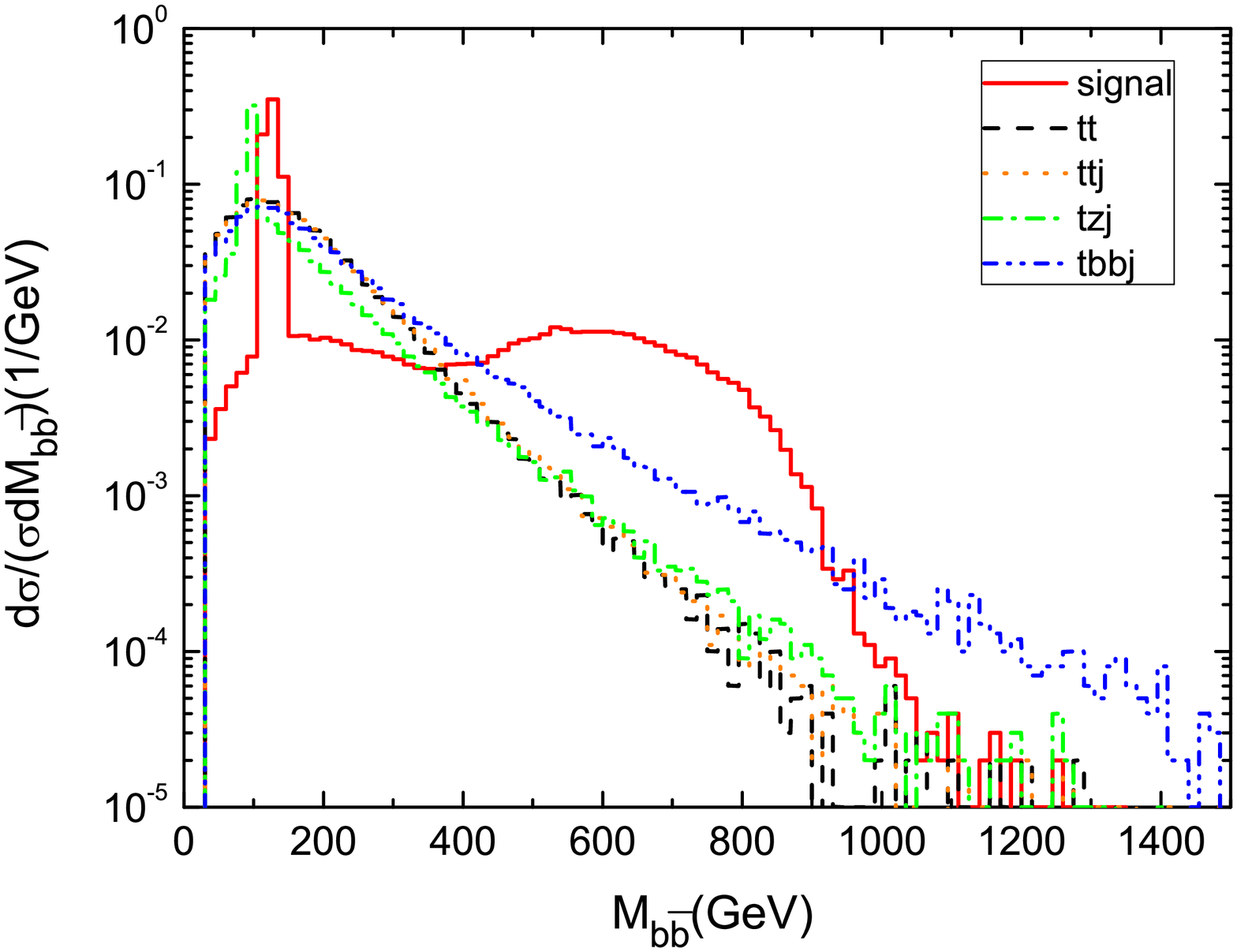}}\hspace{-1.2cm}
\scalebox{0.31}{\epsfig{file=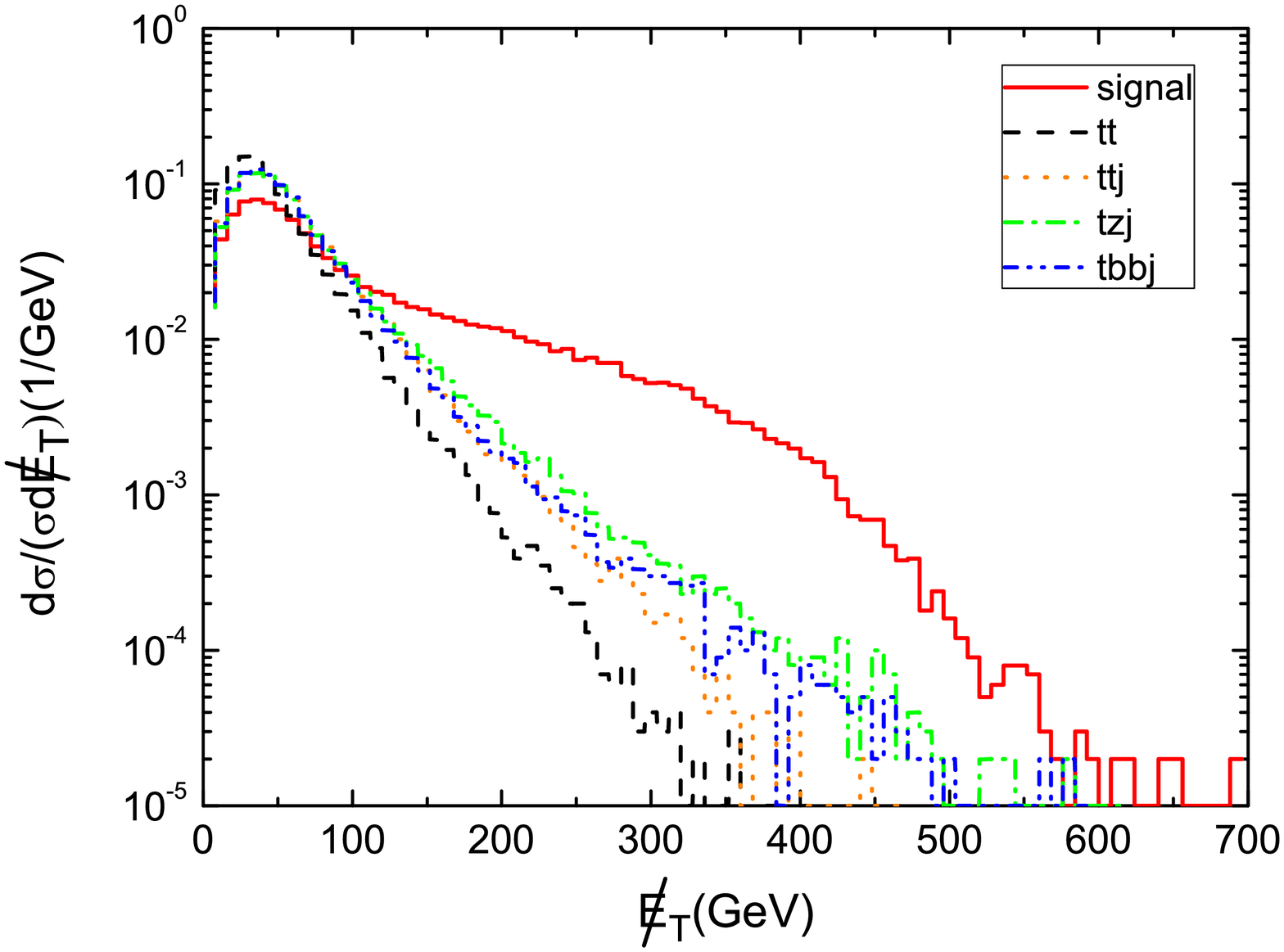}}\vspace{-0.5cm} \caption{The
normalised distributions of $H_{T}, \eta_{j}, M_{b\bar{b}}, \met$
after the basic cuts in the signal and backgrounds at 14 TeV LHC for
$f=700$ GeV, $R=1.5$.}\label{mm}
\end{figure}

After basic cuts, the signal is overwhelmed by the backgrounds. In
order to reduce the contributions of the backgrounds and enhance the
signal contribution, some additional cuts are required and some
other kinematic distributions are needed. As an example, we display
the normalised distributions of $H_{T}, \eta_{j}, M_{b\bar{b}},
\met$ in the signal and backgrounds at 14 TeV LHC for $f=700$ GeV,
$R=1.5$ in Fig.\ref{mm}, where $H_{T}(= \sum_{\text{hadronic
particles}} \big|\big| \vec p_T \big| \big|)$ is the total
transverse hadronic energy, $\eta_{j}$ is pseudorapidity of the
leading jet, $M_{b\bar{b}}$ is the invariant mass of the two
$b$-jets from the Higgs boson decay and $\met(= \big|\big|
\sum_{\text{visible particles}} \vec p_T \big|\big|)$ is the missing
transverse energy.

\begin{table}[ht!]
\fontsize{12pt}{8pt}\selectfont

\caption{Cutflow of the cross sections for the signal and
backgrounds at 14 TeV LHC on the benchmark points [top-left:($f=700$
GeV, $R=1$); top-right:($f=700$ GeV, $R=1.5)$; bottom-left:($f=1000$
GeV, $R=1$); bottom-right:($f=1000$ GeV, $R=1.5$)]. All the
conjugate processes of the signal and backgrounds have been
included.\label{cutflow}}
\begin{center}
\newcolumntype{C}[1]{>{\centering\let\newline\\\arraybackslash\hspace{0pt}}m{#1}}
{\renewcommand{\arraystretch}{1.5}
\begin{tabular}{ C{0.3cm} C{0.3cm} C{0.cm} |C{2.3cm}|C{1.2cm}C{1.2cm} C{1.3cm} C{1.3cm}|C{2cm}| C{2.5cm}}
\cline{1-9} \hline
&\multicolumn{2}{c|}{\multirow{3}{*}{Cuts}}&\multicolumn{5}{c|}{$\sigma$(fb)}
&\multicolumn{1}{c|}{\multirow{2}{*}{$\frac{S}{\sqrt{S+B}}$}}&\multicolumn{1}{c}{\multirow{2}{*}{$\frac{S}{B}$}}\\\cline{4-8}
&&&\multicolumn{1}{c|}{Signal} &\multicolumn{4}{c|}{Backgrounds}&&
\\\cline{4-10}
&&&\multicolumn{1}{c|}{$thj$}
&$t\overline{t}$&$t\overline{t}j$&$tZj$&$tb\bar{b}j$&300fb$^{-1}$&\%\\\cline{1-10}
\multicolumn{2}{c}{\multirow{2}{*}{Basic cuts}}&&1.12(1.34)
&702.7&648.7&1.68&2.82&0.53(0.63)&0.083(0.099)\\
&&&0.72(0.74) &702.7&648.7&1.68&2.82&0.34(0.35)&0.053(0.055)
\\\hline\multicolumn{2}{c}{\multirow{1}{*}{ $H_{T}>530$GeV }}&
&0.45(0.65)&15.27&69.58& 0.16& 0.25&0.85(1.22)&0.53(0.77)
\\\multicolumn{2}{c}{\multirow{1}{*}{ $H_{T}>600$GeV }}&
&0.10(0.11)&7.22&40.24& 0.10& 0.16&0.25(0.29)&0.21(0.24)\\\hline
\multicolumn{2}{ c }{\multirow{2}{*}{ $ |\eta_j|>2$}}&
&0.36(0.54)&0.6&11.13&0.074&0.13&1.78(2.65)&3.0(4.5)\\ &&
&0.07(0.084)&0.23&5.98&0.042&0.076&0.48(0.57)&1.1(1.3)\\\hline
\multicolumn{2}{c}{\multirow{2}{*}{$ |M_{b\bar{b}}-m_h|<
15\text{GeV}$}}&&0.12(0.18)&0.004&0.99&0.0029&0.0034&1.96(2.87)&12(18)\\&&&0.023(0.028)&0.&0.47&0.0015&0.0019&0.57(0.68)&4.9(5.9)\\
\hline\multicolumn{2}{c}{\multirow{1}{*}{$\met
>100\text{GeV}$}}&&0.078(0.12)&0.&0.23&0.001&0.001&2.4(3.6)&33.6(53.5)\\\multicolumn{2}{c}{\multirow{1}{*}{$\met
>180\text{GeV}$}}&&0.01(0.013)&0.&0.036&0.0002&0.0002&0.81(1.02)&27.8(36.1)\\\hline
\end{tabular}}
\end{center}
\end{table}
Firstly, we can see that there is a bulge in the $H_{T}$
distribution of the signal, which arises from the resonance effect
of the top partner $T_{+}$ and this effect also appears in some
other distributions. So we require the events to satisfy $H_{T}>530$
GeV to isolate the signal and find all the backgrounds are
suppressed effectively. After this cut, the backgrounds are still
much larger than the signal, especially the two reducible
backgrounds $t\bar{t}$ and $t\bar{t}j$. According to the $\eta_{j}$
distribution, we can see that most events of $t\bar{t}$ and
$t\bar{t}j$ have a leading jet in the central region, which differs
significantly from the signal, so we apply the cut $|\eta_j|>2$ to
further suppress the $t\bar{t}$ and $t\bar{t}j$ backgrounds.

Another effective cut which can suppress the backgrounds is the
invariant mass cut on the two $b$-jets from the Higgs boson decay.
The $b$ quark from the top decay can be selected with high purity by
choosing the smallest invariant masses $M_{bl}$ of each $b$-jet and
the lepton among the three combinations\cite{mbl}. The other two $b$
quarks are then considered the $b$ quarks coming from the Higgs
boson decay. We find that the signal peak of $M_{b\bar{b}}$ is more
narrow than those of the backgrounds, so we use the cut
$|M_{b\bar{b}}-m_h|< 15$ GeV to enhance the observability of the
signal. Besides, we apply the cut $\met>100$ GeV to further isolate
the signal and find that the $t\bar{t}j$ background is suppressed
effectively. After all cuts above, the background is dominated by
$t\bar{t}j$ completely due to an extra hard jet with $t\bar{t}$.

For easy reading we summarize the cut-flow cross sections of the
signal and backgrounds for 14 TeV LHC in Table.\ref{cutflow}. For
comparison, we chose four sets of benchmark points, that is ($f=700$
GeV, $R=1$),($f=700$ GeV, $R=1.5)$,($f=1000$ GeV, $R=1$) and
($f=1000$ GeV, $R=1.5$), they are arranged in the
top-left(top-right) and bottom-left(bottom-right) part of one cell,
respectively. Due to the resonance effect of the top partner
$T_{+}$, the values of the cuts, mainly the $H_{T}$ and $\met$ cuts,
need to be changed with the top partner mass in order to better
suppress the backgrounds. From the Table.\ref{mtp}, we can see that
the larger $m_{T_{+}}$ will correspond to the larger $H_{T}$ and
$\met$ cuts. Because of the approximate $m_{T_{+}}$, the same values
of $H_{T}$ (or $\met$) cuts are taken for the benchmark points
($f=700$ GeV, $R=1$) and ($f=700$ GeV, $R=1.5)$ in our simulation,
and so are the benchmark points ($f=1000$ GeV, $R=1$) and ($f=1000$
GeV, $R=1.5$).

\begin{table}[h]
\begin{center}
{\small
\begin{tabular}{r|c|c|c}
\hline \hline
 {Benchmark point  \, \, } & {$m_{T_{+}}$(GeV)} & {$H_{T}-$cut(GeV)}& {$\met-$cut(GeV)}\\[0.04cm]
 \hline
 ($f=700$GeV, $R=1$)& $993.8$         & $>530$   & $>100$    \\[0.02cm]
 ($f=700$ GeV, $R=1.5)$ & $1081.5$        &  $>530$ & $>100$     \\[0.02cm]\hline
 ($f=1000$ GeV, $R=1)$ & $1412.3$        &  $>600$  & $>180$    \\[0.02cm]
 ($f=1000$ GeV, $R=1.5)$ & $1533.5$        &  $>600$  & $>180$    \\[0.02cm]
 \hline \hline
\end{tabular}
} \\[0.1cm]

\end{center}
\caption{The top partner mass $m_{T_{+}}$ and the corresponding
$H_{T}$ and $\met$ cuts for the four sets of benchmark points.}

\label{mtp}
\end{table}

In order to analyze the observability, we calculate the
signal-to-background ratio according to $S/\sqrt{S+B}$ and the
systematic significance $S/B$ for the luminosity $\mathcal
L=300$fb$^{-1}$, where $S$ represents the number of signal events
and $B$ represents the number of background events. From
Table.\ref{cutflow}, we can see that $S/\sqrt{S+B}$ and $S/B$ are
substantially improved by these selected cuts, where the
signal-to-background ratio $S/\sqrt{S+B}$ can reach 3.6$\sigma$ and
systematic significance $S/B$ can reach 53.5\% for $f=700$ GeV,
$R=1.5$. Moreover, it's worth noting that the systematic
significance $S/B$ is enhanced obviously, which will help to draw
the signal from the backgrounds.

\section{Summary}

In the framework of the LHT model, we investigate the $t$-channel
process of $pp\rightarrow thj$ at 8 and 14 TeV LHC. With current
constraints, we find that the cross section can be enhanced
obviously in some parameter space compared to the SM predictions. We
further investigate the observability of $pp\rightarrow thj$ with
decays $t\rightarrow \ell^{+}\nu b$ and $h\rightarrow b\bar{b}$ at
14 TeV LHC for some benchmark points. By performing a simple
parton-level simulation, we find that the observability of the LHT
signal is promising at the high-luminosity LHC.

\section*{Acknowledgement}
We would like to thank Lei Wu for helpful suggestions. This work was
supported by the National Natural Science Foundation of China
(NNSFC) under grants Nos. 11405047, 11305049, by the China
Postdoctoral Science Foundation under Grant No. 2014M561987 and the
Joint Funds of the National Natural Science Foundation of China
(U1404113).

\end{document}